\documentstyle[12pt,aasms4]{article}

\begin{document}

\title{
EFFICIENCY OF MOLECULAR HYDROGEN FORMATION ON SILICATES}
\author{Valerio Pirronello $^{*}$}
\affil{Istituto   di    Fisica, 
Universita' di Catania, Catania,  Sicily,  Italy 
}
\author
{Ofer Biham}
\affil{Racah Institute of Physics, 
The Hebrew University, 
Jerusalem 91904,
Israel.}
\author
{Chi Liu, Lyiong Shen, and Gianfranco Vidali $^{**}$}
\affil{The Solid State Science and Technology Program 
and the Department of Physics, Syracuse University,
Syracuse, NY 13244-1130}

\begin{abstract}
We report on laboratory measurements of molecular hydrogen formation
and recombination on an olivine slab as a function of surface temperature under
conditions relevant to those encountered in the interstellar
medium.  On
the basis of our experimental evidence, we recognize that there are
two main regimes of H coverage that are of astrophysical importance;
for each of them we provide an expression giving the production rate
of molecular hydrogen in interstellar clouds.

\end{abstract}

\section{Introduction}

Recently, we measured 
(\cite{pirro97})
the synthesis of molecular hydrogen for the first time on a surface of
a solid of astrophysical interest, a slab of natural olivine (a
silicate made of a mixture of ($Mg_2SiO_4$ and $Fe_2SiO_4$)), and in
experimental conditions close to those encountered in interstellar
clouds.  The importance of these measurements comes from the relevance
of molecular hydrogen in the interstellar medium and from the
impossibility to form it at the required rate
(\cite{jura}) 
in gas phase reactions in such low density an environment.

This fundamental problem has been addressed by 
several authors (among them, 
\cite{gould,hollenbach,smoluchowsky,leitch,duley,buch}) 
who reached different and sometimes conflicting conclusions.  Most of
the work up to now is based on the following simple expression of the
production rate $R_{H_2}$(cm$^{-3}$sec$^{-1}$) of molecular hydrogen:
\begin{equation}
R_{H_2}= {1 \over 2}  n_H v_H A  n_g S_H \gamma 
\end{equation}
where $S_H$ the sticking coefficient 
(i.e., the probability that an 
atom hitting the surface remains on it), $n_H$  the number density,
$v_H$ the speed of $H$ atoms in
the gas phase, $A$ the cross-sectional area of the grain, and $n_g$ the
number density of dust grains. $\gamma$ is the probability
that two hydrogen atoms on the surface meet and recombine; 
such a probability is usually taken to be equal or
close to one, based on the assumption that, even at interstellar dust
temperatures, the mobility of $H$ adatoms by tunneling is so high that
they meet and form $H_2$ before having a chance to evaporate.

The few experimental attempts 
(\cite{brackmann,schutte})
were performed in conditions and on samples that were too different from the
interstellar ones to be useful.  Therefore, the
choice to measure $H_2$ formation on a natural silicate is justified
by the necessity to evaluate the rate on a more realistic
surface
(\cite{whiffen}) 
than the one of a model, even if we know that some difficulties do
exist.
For example, it has been suggested that silicates might be
destroyed by the UV radiation field
(\cite{hong,greenberg});
furthermore, silicates in space are probably amorphous 
(\cite{draine})
, while specimens on Earth are polycrystalline, and fluxes in the
laboratory must necessarily be much higher than those in the
ISM. We recognize that the morphology of the sample
could make a difference in the processes here investigated, since we
expect the sticking coefficient to be higher for a crystalline rather
than an amorphous structure, while the  atom mobility should be lower.

In this paper we present a set of results on the measured hydrogen
recombination efficiency as a function of sample temperature and on
the kinetics of the processes involved. Such results can shed new light on the problem of
$H$ recombination on interstellar grains.

\section{Experimental}

The apparatus has been described previously
(\cite{pirro97}); 
here we recall a few essential features. The measurements were carried
out in a ultra-high vacuum (UHV) stainless steel chamber with
operating pressure in the $10^{-10} \sim 10^{-9}$ torr range. The
sample is placed on a copper holder that is attached to a cold
finger. The sample temperature, measured by two thermocouples, can be
lowered to 5 K by circulating liquid He. The
sample, a natural slab of olivine mechanically polished and with a 
shiny appearance, is initially 
cleaned in a methanol bath and, once inside the UHV chamber, is heated
to $\sim$ 200 $^{\circ}C$ prior to each run for further cleaning. A
copper shroud surrounds the copper holder except in front where the
sample is placed.  Atomic hydrogen and deuterium (obtained
dissociating $H_2$ and $D_2$ in two distinct water-cooled
radio frequency sources) are introduced into the chamber
through two separate, triple differentially pumped lines. A quadrupole
mass spectrometer is used to detect $HD$. This arrangement
gives two advantages. First, the background at mass
3 ($HD$) is very low in the UHV chamber. Second, $H$ and $D$ atoms
encounter each other for the very first time on the cold target
surface and no undissociated $H_2$ or $D_2$ coming from
the atomic sources could be confused with molecules synthesized on the
cold silicate surface.

The following procedure was used. First, we measured the intensity of
the two well collimated beams of $H$ and $D$ atoms entering the
chamber; then, after positioning the quadrupole probe in front of the
target, we measured the amount of $HD$ coming from the surface.  The
$HD$ signal was monitored both during the irradiation of the surface,
$I_{irr}$ (to detect molecules synthesized and promptly
restored to the gas phase), and just after the end of irradiation, $I_{des}$,
during a Thermal Programmed Desorption (TPD) run, in which $HD$ desorbed during
a rapid ($\sim$ 1 K/sec) heat pulse to the sample. The sum, $I_{irr} + I_{des}$,
was divided by the $H$ and $D$ signal, $I_{in}$, measured on the beams entering 
the chamber to get what we call the recombination efficiency $r$=
($I_{irr} + I_{des}$)/$I_{in}$.
The values of $r$ are corrected to take into account that, together
with $HD$, also $H_2$ and $D_2$ molecules are synthesized; thus, $r$
values (with $0<r<1$) represent the {\it total recombination efficiency of molecular
hydrogen}, and is related to $R_{H_2}$ (the recombination rate) by: $ r=2 R_{H_2}/(n_H v_H A n_g)$.

\section{Results}

The large majority of atoms hitting the surface are
reflected back in the gas phase, since the sticking coefficient is
very low
(\cite{pirro97,lin}). 
These atoms are pumped away, except for some that form $HD$ on the
chamber walls and contribute to the increase in the background
signal. Depending on their mobility, some of those that become
adsorbed on the surface remain trapped in deep adsorption sites, while
others migrate and form $HD$. We conservatively
assume that the difference between the background signal during and
before irradiation is due to $HD$ that has formed on the surface and
has been promptly released in the gas phase.  For this reason, this
yield obtained during irradiation has to be considered an
upper limit.  At the lower end ($\sim 5-7 K)$ of the sample
temperature range, this latter yield contributes only a
few $\%$ to the signal due to $HD$ that is formed on the surface.

In Fig. 1 we plot $r$ (the recombination effciency) at
different irradiation temperatures of the sample.  For each temperature, we
plot the contribution to $r$ measured during irradiation only (triangles), the
contribution coming from the thermal
desorption run (empty squares), and the sum of the two contributions
(filled squares).

Fig. 1 shows that the
contribution to $r$ from hydrogen stuck 
on the surface after irradiation decreases with temperature (since the
residence time decreases with $T$), while the signal during
irradiation increases with temperature, but the cumulative effect is
to cause an overall reduction in molecular hydrogen production.

As discussed in the previous section, the recombination efficiency
during irradiation is evaluated under the very conservative assumption
that the signal during irradiation is to be ascribed solely to
molecules coming from the sample. Furthermore, toward the higher end
of the temperature range, the signal during the TPD run
becomes very small, and the real desorption yield could be smaller
than reported. For these reasons, we think that the
overall recombination efficiency is closer to the lower curve (empty
squares) than to the upper one (filled squares). In our previous paper
(\cite{pirro97}),
the recombination efficiency quoted at 10$\sim$15 K refers to this
lower estimate.

The values and trend as a function of temperature of
$r$  obtained here (see Fig.1) are intermediate between that of Hollenbach
and Salpeter 
(\cite{hollenbach}),
Leicht-Devlin and Williams
(\cite{leitch})
and Buch and Zhang
(\cite{buch})
on one  side, and of Smoluchowski 
(\cite{smoluchowsky})
on the other, but clearly closer to the calculated value of the first
group of authors. In Hollenbach and Salpeter's model $S \sim$0.3,
$\gamma =1$; thus, in that case $r$ would be equal to 0.3 (temperature independent).  
Smoluchowski 's
approach takes into account the amorphous structure of grains and
employs a fully quantum mechanical approach. In that case $r$ is
several orders of magnitude lower and strongly temperature
dependent. Obviously, it would be of great value to carry out
experiments on amorphous samples.

To reach a better understanding of how these results apply to
interstellar environments, we carefully analyzed desorption
spectra.  Desorption dynamics, in fact, can be related in a
straightforward manner to specific microscopic processes occurring on
the surface.  In Fig. 2 a series of TPD runs is shown after
irradiation of $H$ and $D$ at the lowest sample temperature and for
irradiation times from 5 $sec$ to 480 $sec$, where the
saturation coverage is reached at around 60 $min$ of irradiation (see
\cite{pirro97}). 

The analysis of the desorption kinetics shown in the spectra of Fig. 2 can 
be carried out according to the Polanyi-Wigner equation 
(\cite{readhead}):
\begin{equation}
- d{\it N(t)}/dt = {\nu}_{\alpha} N(t)^{\alpha} exp(-E_{des}/k_B T)
\end{equation}
where $ dN(t)/dt$ is the rate of desorption,
$\alpha$ is the so-called reaction order,
$\nu_{\alpha}$ is a frequency factor,
${\it N(t)}$ is number of adsorbed atoms or molecules,
$E_{des}$ is the energy barrier for desorption,
$k_B$ is the Boltzmann constant, and
$T$ is the absolute temperature.

In such a simple Arrhenius-type equation 
the coverage dependence is entirely contained
in the $N$-term, the pre-exponential term represents the frequency
(number of times per second) the adsorbate attempts to overcome the
barrier.  The exponential term represents the fraction of attempts
with the required minimum energy.  An inspection of Eq. (2) shows
that: when $\alpha$ = 0, the desorption rate is independent of
coverage (such as in the case where the reaction is the desorption
from a multilayer adsorbate and desorption spectra show a common low
temperature edge); when $\alpha$ = 1, the desorption rate is
proportional to the surface coverage (for example, when the reaction is
just the desorption from one layer of already formed molecules, the
peak position in desorption spectra doesn't change with coverage and
peak shapes are asymmetric due to the sudden depletion of the
adsorbate); when $\alpha$ = 2, the desorption rate is proportional to
the square of the adsorbate density (as when the reaction occurs
between two adsorbed species that become mobile on the surface and
react with each other before desorbing; the desorption spectra have a
common high-temperature tail and the position of the maximum shifts as
a function of coverage as it can be seen  by
taking the derivative of Eq.(2)
(\cite{weinberg}). 

In the bottom part of the Fig. 2, at the lowest coverage the desorption
kinetics is clearly of the second order. Using this analysis for the curves at
the bottom of Fig.2, we get the desorption energy within $\pm 0.1$
meV. Due to uncertainty in the saturation coverage, we quote $E_{des}$
=25.5$\pm$0.6 meV.

At intermediate coverage,  desorption curves  have a much reduced shift in the peak positions
but they are still symmetric (see Fig. 2 in
\cite{pirro97}
). At higher coverages, but below monolayer
coverage, peak shapes have characteristics of first order desorption
kinetics (top part of Fig. 2), which is typically characterized by the
same peak position (for increasing coverage) with a strong asymmetry
between the low vs. high temperature sides.

\section{Discussion}

Our set of experiments, carried out at temperatures of 5$\sim$7 K and
up, gives convincing evidence of the fact that the formation of
molecular hydrogen on a surface is a thermally activated process
characterized by the existence of a threshold temperature that will
likely depend on the nature (i.e, chemical composition and structure) of
the grain surface.

The fact that in the low coverage regime the desorption is of the
second order means that at low surface temperature atoms accomodate
on the surface and remain localized as atoms without recombining. If
they recombined, desorption would be of the first order. In turn, this
implies that tunneling at low temperature is not sufficient by itself
to assure the mobility of hydrogen adatoms. Such mobility is required
for hydrogen atoms in order to scan all the adsorption sites on the
grain and encounter each other, according to the model of Hollenbach
and Salpeter.

According to the experimental evidence on desorption kinetics, 
and in agreement with the analysis of peaks performed according to Eq.(2), we
propose that, in interstellar clouds, whenever the temperature of
grains is such as to maintain a low coverage regime, the appropriate
expression for the production rate $R_{H_2} (cm^{-3} sec^{-1})$ of molecular
hydrogen , should be:
\begin{equation}
 R_{H_2}= {1 \over 2} (S_H n_H v_H A t_H)^2 n_g \tilde{N}^{-2} \nu f(T,a,
\delta E) {\gamma \prime},
\end{equation}
\noindent
The term squared in Eq. (3) represents the total number of $H$ atoms
on the grain surface and the $\tilde{N}^{-2} \nu f(T,a, \delta E) $
describes the mobility of adatoms due both to thermal activation and
tunneling, two processes that can be in competition with or help each
other. The cross-sectional area $A$ can be written as $N^{2/D}
\sigma$, where 
$N$ is the number of sites, $\sigma$ the area of each adsorption site,
and $D$ is the fractal dimension (likely to be from $2$ to $\sim 2.6$ for
many non-porous carbon and silicon-bearing solids; see,
\cite{avnir});
$t_H = {\nu_H}^{-1} exp(D_H/k_B T)$ is the residence time of $H$ atoms
($\nu_H$ is a characteristic frequency of an H atom in the energy
level $D_H$ inside the adsorption potential well), $\tilde{N}$ is the
average number of sites between two adsorbed $H$ atoms ($\tilde{N}$ is
squared to take into account that adatoms perform random walks), $\nu$
an inverse characteristic time related to atom diffusion, $T$ is the
temperature, $\delta E$ and $a$ are energy and width parameters for
tunneling.  The function $f$ might be obtained, in analogy to  
electric conduction
in non-crystalline materials, by the ``variable range
hopping'' theory: $f$ = $f_0$ $ exp(-B/k_B T^{1/3})$
(\cite{mott}). $\gamma \prime$ is 
the probability that two H adatoms recombine upon encountering.

An additional process to be considered in interstellar environments is the competition between $H$ 
atoms and already formed $H_2$ molecules (coming from the gas phase)
to occupy the available adsorption sites on the grain. In this case, the number of
available adsorption sites on the grain is $(1-S_{H_2}
n_{H_2} v_{H_2} N \sigma t_{H_2})$.

The top panel of Fig.2 shows that at higher coverage, but still lower than one layer,
the shape of
the desorption peaks assume the characteristics of first order
desorption. The straightforward interpretation is that when the
coverage is high enough (close to one monolayer of $H$ adatoms),
molecular hydrogen that is formed at low temperature doesn't leave the
surface until the temperature is raised.  By the way, accepting this
scenario implies that, in astrophysical environments, the release
of 4.5 eV in the recombination reaction doesn't necessarily restore
promptly $H_2$ in the gas phase.  Another possible interpretation is
that adsorbed H atoms are so close to each other that, when mobility
is increased by raising the temperature, a high fraction of them
recombine and are released in the gas phase with a second order
desorption kinetics that is so fast that it mimics a first order one.
 
In this coverage range, the rate limiting process is not diffusion
because $H$ atoms are separated by short distances, and in agreement
with the first order desorption kinetics observed in our experiments, 
and in agreement with Eq.(2),
the appropriate way to calculate the $H_2$ production rate $R_{H_2}
(cm^{-3} sec^{-1})$ in clouds is given by the expression linear in
N(t):
\begin{equation}
R_H{_2}={1 \over 2} (S_H v_H n_H A t_H)
n_g \nu f(T, a, \delta) \tilde{N}^{-1} \gamma \prime
\end{equation}
where the term $\tilde{N}$ doesn't appear  squared as in Eq.(2)
because $H$ atoms don't have  enough room for a random walk.

The results of Fig. 1, 
show a decline with T of a $\sim$ factor ten from 5 K to
15 K. This cannot be described by Eq.(1) (no T dependence), nor Eqs.(3) and 
(4) (far steeper T dependence with $\gamma \prime =1$). 
We suggest that $\gamma \prime$ (Eqs. (3) 
and (4)) might be different from 1 and temperature dependent. 
This could be the case if there is a small activation barrier $E_{\gamma \prime}$ for 
recombination. Take $D_H$$\sim$15 meV (obtained by assuming that $D_H$/$D_{H_2}$ $\sim $ $ {3 \over 4}$ 
is the same for olivine and graphite and that the desorption energy $E_{des} \sim$ adsorption energy); then the temperature dependence of $r$ in Fig. 1 is 
reproduced with $E_{\gamma \prime}$ $\sim$18 meV and $\sim$5 meV if Eq.(3) 
and (4) are used, respectively (We assumed a barrier for diffusion 
$\sim$0.3 $D_H$  
 and have taken the usual corrections for isotopes).

Several could be the implications of these results for both the
astrochemistry and general understanding of the dust component of
interstellar clouds. For instance, as a consequence of the increase of
the exposed grain surface (see 
\cite{pirro97}
), due to the necessity to reconcile our experimental data with
astronomical observations
(\cite{jura}), 
rates of other surface reactions will have to be evaluated to reassess
their importance at the light of the increased surface area of the
grain.
Such an increment in the area will be probably due to an amorphous or
porous structure of the grain, whose surface will be characterized by
a high degree of roughness, the presence of voids, etc. As a major
consequence, optical properties of grains could be significantly
affected.

Certainly more experimental effort in the study of reaction rates on
amorphous surfaces is necessary to further clarify the implications of
these measurements.

\vskip 0.3 true in \noindent{\large Acknowledgments}

\acknowledgments
Support from NASA-Astrophysics Division and the Italian National
Research Council is gratefully
acknowledged. We thank Dr. P.Plescia of the CNR - Rome for providing the
olivine sample and  ENI of Rochester, N.Y. for donating the RF
splitter. Helpful discussions with Prof. Eric Schiff of S.U. are
gratefully acknowledged. We also thank the anonymous referee for key suggestions.
\vskip 0.3 true in \noindent
$^{*}$ E-mail: vpirrone@cdc.unict.it

\noindent
$^{**}$ To whom correspondence should be addressed; e-mail:
gvidali@syr.edu

\vspace{0.5in}

\newpage
\figcaption{Recombination efficiency $r$ of hydrogen on an olivine
slab as a function of temperature: computed during irradiation time
(triangles), from the thermal desorption run (empty squares), and
sum of the two contributions above (filled squares). 
Irradiation time: $\sim 1 min.$ Lines are
guides to the eye.}

\figcaption{Desorption rate of $HD$ during thermal desorption runs
from an olivine slab at T$\sim 6$ K. Bottom panel: after irradiation
for (bottom to top) 0.07, 0.10, 0.25, 0.55 minutes; top panel: after
irradiation for (bottom to top) 2.0, 5.5 and 8.0 minutes.}

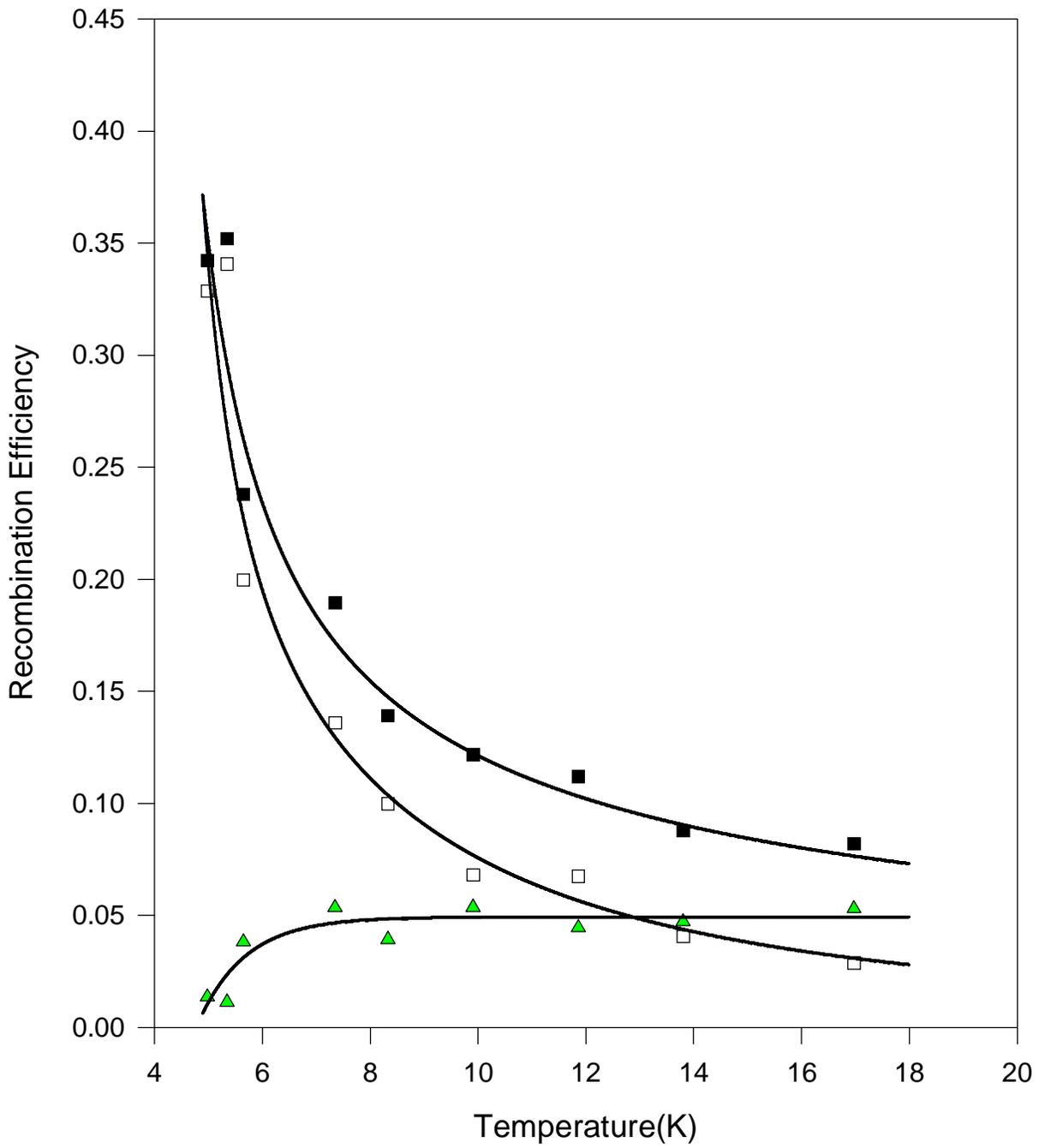
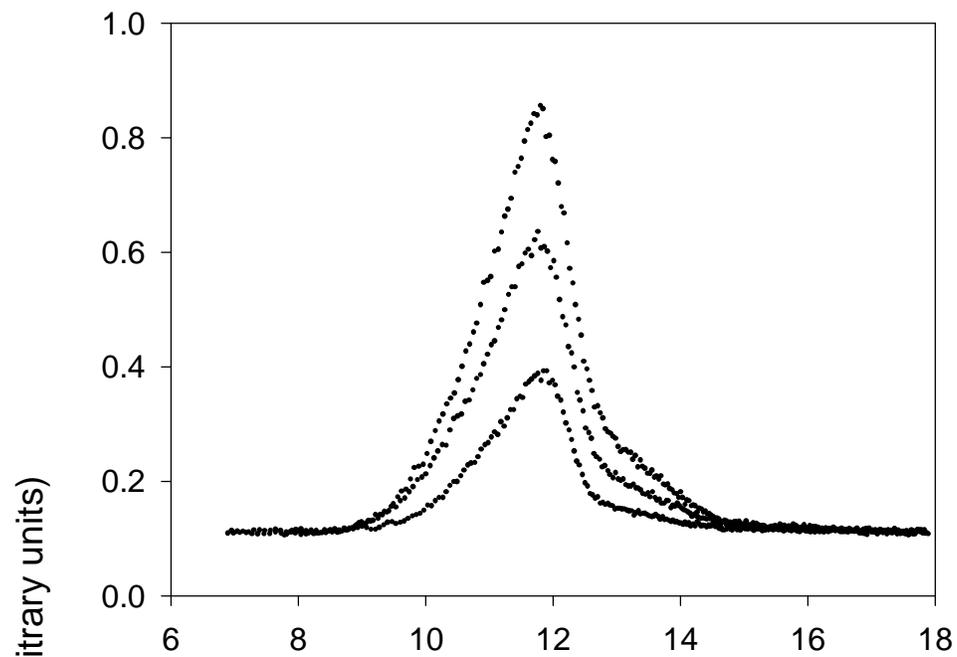
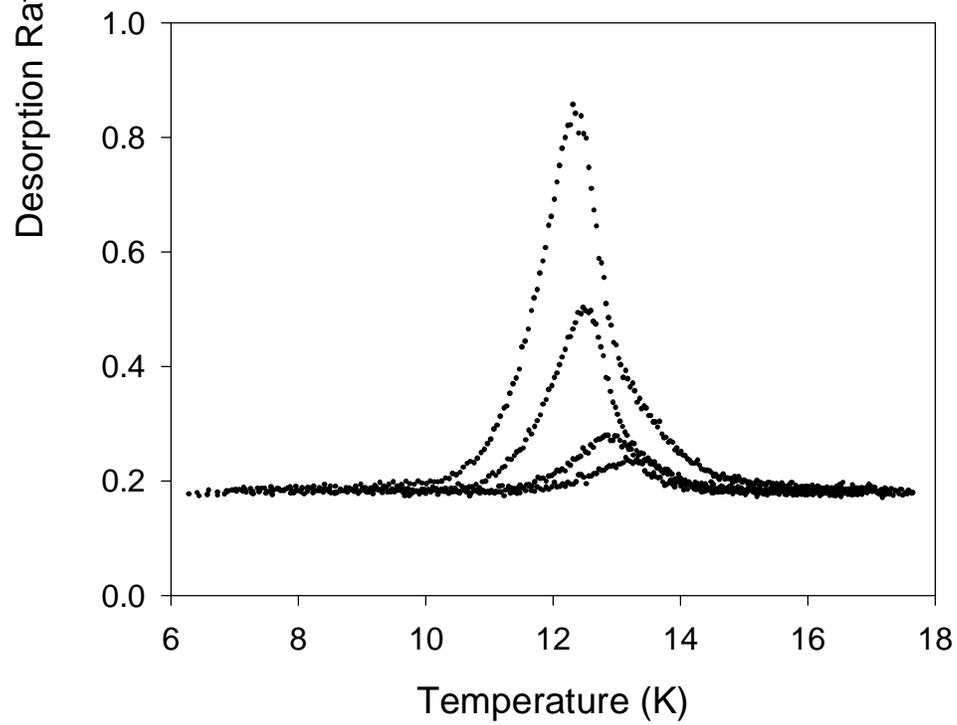
\end{document}